\documentclass[9pt,twocolumn,twoside]{osajnl}

\journal{ol} 

\setboolean{shortarticle}{true}

\usepackage{multirow}

\newcommand{\fref}[1]{Figure~\ref{#1}}
\newcommand{\tref}[1]{Table~\ref{#1}}
\newcommand{\eref}[1]{Equation~\ref{#1}}
\newcommand{\cref}[1]{Chapter~\ref{#1}}

\title{Structure-Aware Parametric Representations for Time-Resolved Light Transport}
\doi{\url{https://dx.doi.org/10.1364/OL.465316}}

\author[1,*]{Diego Royo}
\author[2,*]{Zesheng Huang}
\author[2,**]{Yun Liang}
\author[2]{Boyan Song}
\author[1]{Adolfo Muñoz}
\author[1]{Diego Gutierrez}
\author[1]{Julio Marco}

\affil[1]{Universidad de Zaragoza---I3A, Zaragoza, Spain}
\affil[2]{South China Agricultural University, Guangzhou, China}

\affil[*]{Joint first authors}
\affil[**]{Corresponding author: yliang@scau.edu.cn}

\begin{abstract}
Time-resolved illumination provides rich spatio-temporal information for applications such as accurate depth sensing or hidden geometry reconstruction, \new{becoming a useful asset for prototyping and as input for data-driven approaches}. However, time-resolved illumination measurements are high-dimensional and have a low signal-to-noise ratio, hampering their applicability in real scenarios. We propose a novel method to compactly represent time-resolved illumination using mixtures of exponentially-modified Gaussians that are robust to noise and preserve structural information. Our method yields representations two orders of magnitude smaller than discretized data, \new{providing consistent results in applications such as hidden scene reconstruction and depth estimation, and quantitative improvements over previous approaches.}
\end{abstract}

\setboolean{displaycopyright}{true}

\begin{document}

\newcommand{\new}[1]{#1}

\maketitle

\graphicspath{ {./figures/} }


\begin{figure*}[!t]
	\centering
	\includegraphics[width=\linewidth]{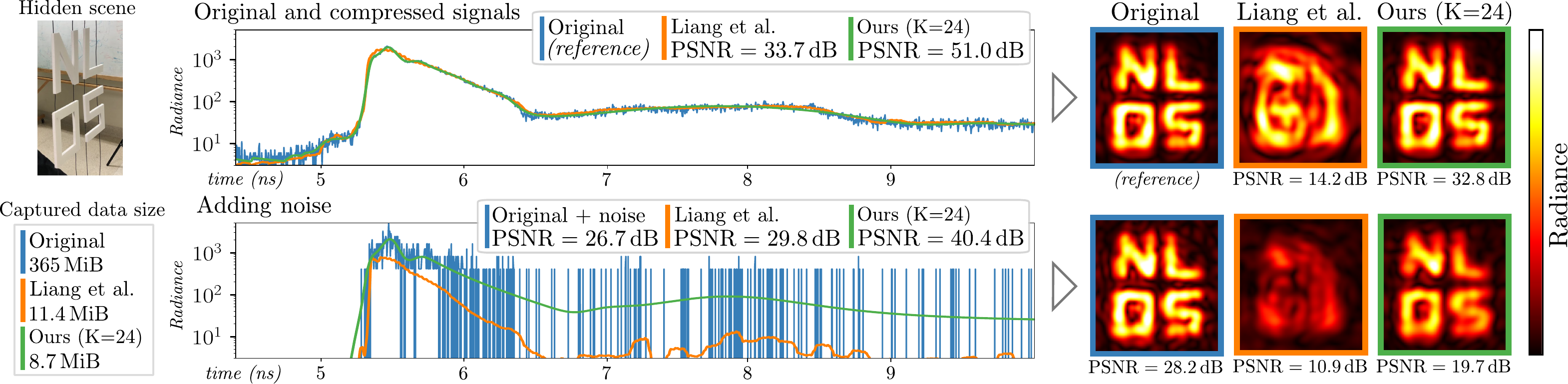}
	\caption{Application of our method to non-line-of-sight (NLOS) data for a real scene. Top: original captured signal \new{histogram}. Bottom: adding noise \new{to the histogram} simulating an exposure time 200 times shorter. The \new{four letters in the scene are} then reconstructed following the work by Liu et al. \cite{liu2019non}. On the right, we show the results using the original captured and compressed signals using the autoencoder network proposed by Liang et al. \cite{liang2020compression}, and our method.}
	\label{fig:nlos-reconstruction}
\end{figure*}
Transient imaging methods analyze time-resolved light transport at very high temporal resolutions, with applications such as reconstruction of hidden geometry \cite{Lindell2019wave,liu2019non,Xin2019theory}, object detection through scattering media \cite{Heide2014}, or material classification \cite{su2016material}, with promising advances over the recent years \cite{jarabo2017recent}. Current capture methods combine ultra-fast lasers with sensors such as Single Photon Avalanche Diode (SPAD) arrays \cite{renna2020fast, nam2021low}, providing dense spatial scanning and picosecond temporal resolution that yield rich spatio-temporal information of the captured scene.

These capture setups, however, introduce several limitations. 
First, analyzing the spatio-temporal structure of indirect illumination is fundamental in many transient imaging applications, but \new{multiple-scattered} light can become too attenuated when reaching the sensor. Consequently, imaging applications may be subject to measurements with a low signal-to-noise ratio (SNR), which can decrease their performance. 
%
%
Second, dense temporal and spatial resolutions of the measurement space result in high memory and bandwidth requirements (with datasets of tens or hundreds of gigabytes) \cite{marco2021NLOSvLTM}, which can become a bottleneck in light transport analysis and application design.

In this work, we provide a method for lightweight representations of time-resolved illumination with a three-fold benefit: the representation space is up to two orders of magnitude smaller than source data and two times smaller than previous compression approaches, it is robust to noise, and it preserves structural information fundamental in transient imaging applications. We rely on mixtures of exponentially-modified Gaussian (EMG) distributions and design an optimization procedure that accounts for spatial gradients to preserve structural information.

Several previous works propose alternative representations of time-resolved light pulses with different goals. Note that none of these methods is structure-aware, as they only use temporal information from single points in the scene.
Peters et al. \cite{Peters2015} use the Pisarenko and maximum entropy spectral estimates to reconstruct transient pulses and improve data quality by removing multipath interference in range imaging.
For these same purposes, Kadambi et al. \cite{Kadambi2013} recover per-pixel sparse time profiles expressed as a sequence of impulses.
Other works are based on linear inverse problems \cite{Heide2014diffuse} solved by numerical optimization, and frequency-domain reconstructions \cite{Lin2014} which introduce Fourier analysis to reduce systematic errors.
%
%
Closer to our representation space, Wu et al. \cite{Wu2014} analyze direct and indirect illumination components by representing light pulses as a combination of one Gaussian and one exponentially-modified Gaussian.
Heide et al. \cite{Heide2014} recover time-resolved illumination in turbid media from correlation-based sensors by sparsely encoding light with EMG distributions.
%

Directly related to our goal, Liang et al. \cite{liang2020compression} recently introduced feature-based compact representations of time-resolved illumination using deep encoder-decoder neural architectures. However, their approach is biased towards line-of-sight training data and fails to preserve structural information that is fundamental in modern transient imaging applications. As shown in \fref{fig:nlos-reconstruction}, our method yields representations of \new{previously captured} transient illumination \new{histograms} with higher quality and that preserve structural information, providing significantly better performance on hidden geometry reconstructions than feature-based methods, even with a higher compression ratio.

Inspired by previous works \cite{Wu2014,Heide2014}, we propose to use exponentially modified Gaussian (EMG) distributions to compactly represent time-resolved illumination. Consider a transient camera which adds a third temporal dimension $T$ for an image $I$ with size $W \times H$, i.e. $I[i, j, t] \in \mathbb{R}^{W \times H \times T}$. A time-resolved pixel $I_\textbf{p}[t] \in \mathbb{R}^{1 \times 1 \times T}$ at $\textbf{p} = \{i, j\}$ represents the accumulation of light paths with a timestamp $t$ traveling from the light sources to the sensor pixel $\textbf{p}$ after being scattered through the scene elements. Time-resolved illumination typically has the temporal shape of aggregated radiance pulses with exponential decay. This behaviour stems from multi-bounce convolutions of the source illumination pulse with the scene geometry. 
EMG distributions model a Gaussian with a parameterized exponential decay, and therefore arise as a convenient function to model the physical behaviour of different illumination bounces in time \cite{Wu2014,Heide2014}. We, therefore, propose to represent the response of a pulsed source measured at an ultra-fast sensor pixel by aggregating EMG distributions in a mixture model. 
\new{We analyze the benefits of our method on real data captured on non-line-of-sight (NLOS) configurations \cite{liu2019non}, and simulated datasets \cite{MarcoSIGA2017DeepToF,galindo19-NLOSDataset}, which have proved to faithfully represent data captured with real hardware. We demonstrate the consistency of our representation in applications such as NLOS reconstruction \cite{liu2019non}, and line-of-sight (LOS) depth estimation based on amplitude-modulated continuous-wave time-of-flight (AMCW ToF) sensors \cite{MarcoSIGA2017DeepToF}.} 


\paragraph{Exponentially-modified Gaussians}
An exponentially-modified Gaussian (EMG) distribution is defined as
\begin{align}
\operatorname{EMG}(t; h, \mu, \sigma, \tau )=\frac{h\sigma}{\tau} \sqrt{\frac{\pi}{2}}\exp \left( \frac {1}{2} \left( \frac {\sigma}{\tau} \right)^2 - \frac {t-\mu}{\tau} \right) \nonumber \\
* \operatorname{erfc} \left( \frac {1}{\sqrt{2}}\  \left( \frac {\sigma}{\tau}  - \frac {t-\mu}{\sigma} \right) \right ),
\label{eq:EMG}
\end{align}
where $h$ controls the pulse amplitude, $\tau$ controls the exponential decay rate, and the mean $\mu$ and standard deviation $\sigma$ control the peak position and width of the Gaussian distribution. The complementary error function, $\text{erfc}(\cdot)$, is defined as
\begin{equation} \label{eq:ERFC}
\operatorname {erfc}(x)=
{\frac  {2}{{\sqrt  {\pi }}}}\int _{x}^{\infty }e^{{-t^{2}}}\,dt.
\end{equation}
In our method we model the aggregation of multi-bounce light paths on a time-resolved single pixel $I_\textbf{p}[t]$ as an EMG mixture model ($\operatorname{EMGMM}$) with $K$ EMGs, denoted as $I_\textbf{p}^{\prime}[t]$,
\begin{align}
{I}_{\textbf{p}}[t] \approx {I}_{\textbf{p}}^{\prime}[t] &  = \operatorname{EMGMM}(t; \boldsymbol{h_{p}}, \boldsymbol{\mu_{p}}, \boldsymbol{\sigma_{p}}, \boldsymbol{\tau_{p}}, K)  \nonumber \\
&= \sum _{k=1}^{K} \operatorname{EMG}(t; h_k, \mu_k, \sigma_k, \tau_k),
\label{eq:EMGMM}
\end{align}
where each $\boldsymbol{a}_p$ term represents a vector of $K$ EMG parameters $a_k$ with $k=1..K$ for the estimation of pixel $\textbf{p}$.

We formulate the estimation of a transient pixel $I_\textbf{p}[t] \approx I_\textbf{p}^\prime[t]$ as an optimization problem which attempts to compute the best EMGMM parameterization $\{\boldsymbol{h_p},\boldsymbol{\mu_p},\boldsymbol{\sigma_p},\boldsymbol{\tau_p}\}$ in order to minimize the pixel loss function $\mathcal{L}_p$:
\begin{align}
\mathop{\arg\min}\limits_{\boldsymbol{h_p},\boldsymbol{\mu_p},\boldsymbol{\sigma_p},\boldsymbol{\tau_p}} \mathcal{L}_p(I_\textbf{p},I_\textbf{p}^{\prime}),
\label{eq:single_pixel_optimization}
\end{align}
where $\mathcal{L}_p$ is lower when both time-resolved illumination distributions are more similar. Since we are working with EMG distributions, we define $\mathcal{L}_p$ based on the Kullback-Leibler Divergence (KLD) \cite{kullback1951information}, a statistical distance metric used to measure the difference of probability distributions. For a pixel $I_\textbf{p}$ and its reconstruction $I_\textbf{p}^{\prime}$, the KLD is defined as
\begin{align}
\displaystyle D_{\text{KL}}(I_\textbf{p}^{\prime} \parallel I_\textbf{p}) = \sum_{t\in T}I_\textbf{p}^{\prime}[t]\log \left({\frac {I_\textbf{p}^{\prime}[t]}{I_\textbf{p}[t]}}\right).
\end{align}
Note that KLD is asymmetric, and also produces low errors where $I_\textbf{p}^{\prime} \approx 0$ and high errors where $I_\textbf{p} \approx 0$. To avoid these extreme cases, we use its asymmetric property to finally construct our pixel loss function $\mathcal{L}_p$:
\begin{align}
  \mathcal{L}_p(I_\textbf{p},I_\textbf{p}^{\prime}) &= \displaystyle D_{\text{KL}}(I_\textbf{p}\parallel I_\textbf{p}^{\prime})+D_{\text{KL}}(I_\textbf{p}^{\prime}\parallel I_\textbf{p})
  \nonumber \\
  &= \sum_{t\in {T}}(I_\textbf{p}[t]-I_\textbf{p}^{\prime}[t])(\log{(I_\textbf{p}[t])} - \log{(I_\textbf{p}^{\prime}[t])}).
  \label{eq:lossfunction}
\end{align}
\begin{figure}[!t]
	\centering
	\includegraphics[width=\columnwidth]{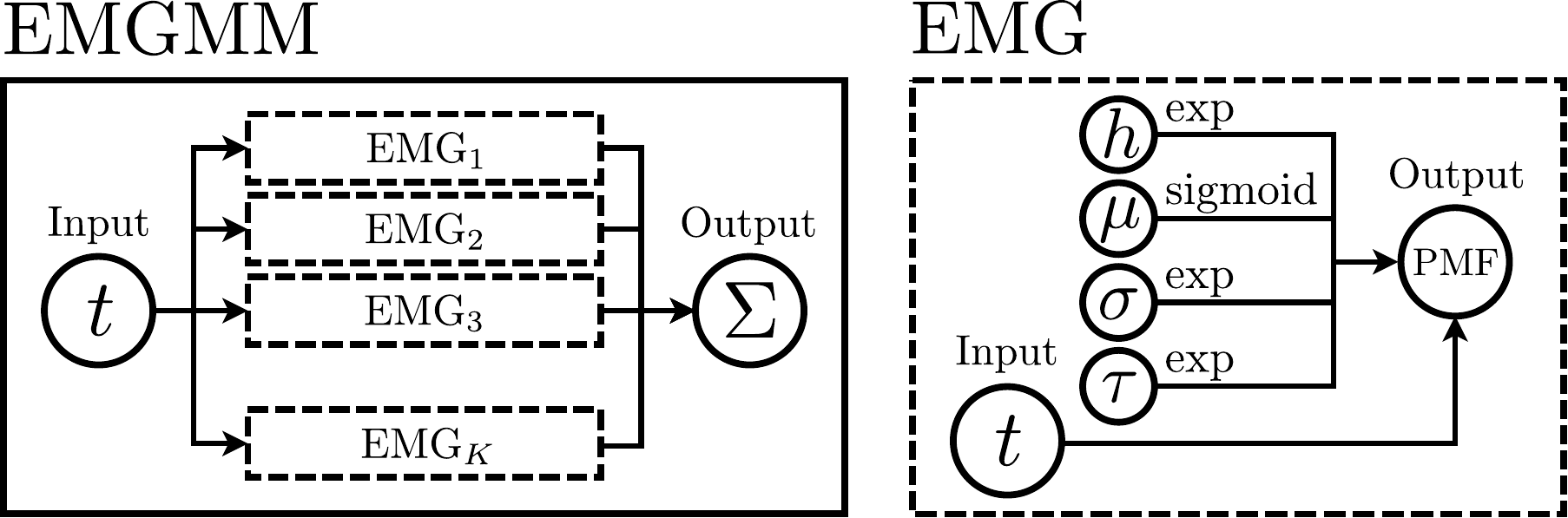}
	\caption{Left: EMGMM with $K$ EMG modules. Right: EMG substructure with four parameters and one input connected to the Probability Mass Function (PMF) node corresponding to \eref{eq:EMG}. We use exponential functions to ensure $h, \sigma, \tau > 0$. As $t \in [0, 1]$ is normalized, a $\text{sigmoid}$ function is applied to $\mu$.}
	\label{fig:model_structure}
 \vspace{-1em}
\end{figure}
Given that $\mathcal{L}_p$ (\eref{eq:lossfunction}) and EMGs (Equations \ref{eq:EMG} and \ref{eq:ERFC}) are differentiable, we use Stochastic Gradient Descent to optimize for the best parameters in an EMG-based differentiable pipeline shown in \fref{fig:model_structure}. For an input time $t$, the pipeline should output the pixel intensity at that time $I_\textbf{p}^{\prime}[t]$.

Other loss functions such as the Mean Square Error (MSE) \cite{liang2020compression} tend to be biased towards large-valued regions in the temporal domain, which may lead the optimization to fall into local minima. Our KLD-based optimization reduces the error more uniformly across the entire temporal domain, shown in \tref{tab:optimization-order} when compared to the MSE-based approach.

For practical reasons, we pre-process the optimization input as follows: we first clip the temporal domain of each pixel $I_\textbf{p}[t]$ to $t\in[t_\text{start}, T]$, where $I_\textbf{p}[t_\text{start}]$ \new{is} the first non-zero value. We then normalize the clipped temporal domain to $t \in [0, 1]$ \new{by} subtracting $t_\text{start}$ and dividing by the resulting length $T' = T - t_\text{start}$. The EMGMM (\eref{eq:EMGMM}) is evaluated in a reparameterized time interval $t\in [0, 1]$, where the reference pixel $I_\textbf{p}[t]$ does not have leading zeroes. We store $t_{start}$, $T'$ along with the $4K$ EMGMM parameters $\boldsymbol{h_p},\boldsymbol{\mu_p},\boldsymbol{\sigma_p},\boldsymbol{\tau_p}$, and revert the clipping and normalization to compute the loss (\eref{eq:lossfunction}), resulting in a compression factor of ${T}/{(4K+2)}$ \new{and optimization runtime within $O(KT)$.}

\paragraph{Reconstructing a transient image} \eref{eq:single_pixel_optimization} defines the optimization scheme to represent a single pixel $I_\textbf{p}[t]$, with $\textbf{p}=\{i,j\}$ using a EMGMM. \new{In} a full transient image $\equiv I[i, j, t]$, neighbouring pixels usually present significant spatio-temporal structure. We propose an optimization methodology to use spatio-temporal information to improve the pixel representation in a transient image with a two-fold benefit. First, reduction of noise in low SNR areas by leveraging noisy information from nearby pixels. Second, preserving the spatial structure of the signal by accounting for the spatial gradient. In particular we use the $N \times N$ window around each pixel $\textbf{p}$, noted as $\mathcal{W}(\textbf{p}, N) \equiv \mathcal{W}$ and defined as all pixels $\textbf{p}^{\prime}$ where $\Vert \textbf{p}^{\prime} - \textbf{p} \Vert_\infty < N/2$. We introduce a spatial gradient loss term $\mathcal{L}_g$ that fosters consistency between neighbouring pixels, reducing the influence of per-pixel noise:
\begin{align}
  \mathcal{L}_g(I_\mathcal{W},I_\mathcal{W}^{\prime}) = \sum_{t \in T} & \Big( \parallel{G_\text{up}(I_\mathcal{W}, t) - G_\text{up}(I_\mathcal{W}^{\prime}, t)}\parallel_2 \nonumber \\
  &+ \parallel{G_\text{down}(I_\mathcal{W}, t) - G_\text{down}(I_\mathcal{W}^{\prime}, t)}\parallel_2 \nonumber \\
  &+ \parallel{G_\text{left}(I_\mathcal{W}, t) - G_\text{left}(I_\mathcal{W}^{\prime}, t)}\parallel_2 \nonumber \\
  &+ \parallel{G_\text{right}(I_\mathcal{W}, t) - G_\text{right}(I_\mathcal{W}^{\prime}, t)}\parallel_2 \Big).
  \label{eq:LossG}
\end{align}
Each gradient $G$ is defined for a neighbourhood $\mathcal{W}$ where each element is computed using its immediate neighbours as
\begin{equation}
	\left\{
	\begin{aligned}
 	&G_\text{up}(I_\mathcal{W}, t) &= I[i, j - 1, t] - I[i, j, t] \\
  	&G_\text{down}(I_\mathcal{W}, t) &= I[i, j + 1, t] - I[i, j, t] \\
  	&G_\text{left}(I_\mathcal{W}, t) &= I[i - 1, j, t] - I[i, j, t] \\
 	&G_\text{right}(I_\mathcal{W}, t) &= I[i + 1, j, t] - I[i, j, t]
	\end{aligned}
	\right.
	, \, \textbf{p} = \{i, j\} \in \mathcal{W},
	\label{eq:gradient_loss}
\end{equation}
adding zero-padding on the image to satisfy the calculation for edge pixels. The final optimization problem is
\begin{align}
\mathop{\arg\min}\limits_{\boldsymbol{h}_\mathcal{W},\boldsymbol{\mu}_\mathcal{W},\boldsymbol{\sigma}_\mathcal{W},\boldsymbol{\tau}_\mathcal{W}}
\mathcal{L}_g(I_\mathcal{W},I_\mathcal{W}^{\prime}) +
\sum_{\textbf{p} \in W}
\mathcal{L}_p(I_\textbf{p},I_\textbf{p}^{\prime}),
\label{eq:final_optimization_problem}
\end{align}
reconstructing $N \times N$ pixels in the image simultaneously. The resulting optimization accounts for multiple EMGMMs as in \fref{fig:model_structure} for each pixel in the neighbourhood, calculating the gradient afterwards. \new{Typical values are $N \in \{3, 5, 7\}$. From our experiments, $N=5$ provides the best tradeoff between performance and time, with little difference from other values.} Also, the data needs to be pre-processed as explained previously. The value of $t_\text{start}$ is obtained as the minimum for all pixels $\textbf{p} \in \mathcal{W}$ based on the first pixel that receives a photon. The compression ratio for the whole neighbourhood is $(N^2 \cdot T) / (N^2\cdot 4K + 2)$ \new{with an execution time of the optimization within $O(N^2KT)$.}
\begin{table}[tb]
\caption{KLD loss $\mathcal{L}_I = \sum_\textbf{p} \mathcal{L}_{p}(I_\textbf{p}, I'_\textbf{p})$ in a region of the \emph{Staircase} NLOS scene \cite{galindo19-NLOSDataset} with $K=64$ for pixel-independent ($\mathcal{L}_{p}$), and \emph{Sliding (S)} and \emph{Random (R)} structure-aware fitting ($\mathcal{L}_{p} + \mathcal{L}_g$). The \emph{Random} fitting is also tested with an MSE loss.}
\label{tab:optimization-order}
\begin{tabular}{l|r|r|r|r|}
\cline{2-5}
 & \multicolumn{3}{c|}{\textbf{KLD loss}} & \multicolumn{1}{c|}{\textbf{MSE loss}} \\ \cline{2-5} 
 & \multicolumn{1}{c|}{$\mathcal{L}_p$} & \textbf{\begin{tabular}[c]{@{}c@{}}$\mathcal{L}_p + \mathcal{L}_g$ \emph{(S)}\end{tabular}} & \textbf{\begin{tabular}[c]{@{}c@{}}$\mathcal{L}_p + \mathcal{L}_g$ \emph{(R)}\end{tabular}} & \textbf{\begin{tabular}[c]{@{}c@{}}$\mathcal{L}_p + \mathcal{L}_g$ \emph{(R)}\end{tabular}} \\ \hline
\multicolumn{1}{|l|}{$\mathcal{L}_I$} & $0.012$ & $0.010$ & $\mathbf{0.008}$ & $0.010$ \\ \hline
\end{tabular}
\end{table}

\paragraph{Initialization}
Since our optimization is based on a Stochastic Gradient Descent, a good initialization is crucial to avoid converging to bad local minima. Light transport in real-world scenes decays exponentially: short, high-energy paths arrive early and have higher temporal frequency while lower-energy, \new{\mbox{multiple-scattered}} paths with longer optical \new{lengths} are smooth and arrive later in time. We use these observations to estimate convenient initial values for the peak $\mu$, width $\sigma$ and decay rate $\tau$ of each of the $K$ EMG modelled pulses. Note that the amplitude $h$ is just a scale factor, so we arbitrarily initialize it as $h=1$.

For each parameter, we initialize values $v_i$ in log space, which provides a 5-10\% error improvement with respect to linear-space initialization for a similar number of epochs. We uniformly sample $\xi_i \in [\log(v_\text{min}),\log({v_\text{max}})]$ and then convert to linear range as $v_i = \exp(\xi_i)$. First, we sample $K$ values for $\mu$ and $\sqrt{K}$ values for $\sigma$ and $\tau$, of which we generate all possible $K$ combinations. The first bounces with smaller $\mu$ are narrower, so they are assigned the smaller $\sigma$ and $\tau$ values. This is done for each of the $K$ values of $\mu$ and $(\sigma, \tau$) pairs. Under this initialization, illumination pulses centered at later timestamps $\mu$ will have a wider support $\sigma$ and a slower decay $\tau$.

To reconstruct an image $I[i, j, t] \in \mathbb{R}^{W \times H \times T}$ the order for which we optimize the pixels is important when using spatial consistency as in \eref{eq:final_optimization_problem}. A first approach would be to slide through the $N \times N$ image pixels while optimizing each pixel neighbourhood. This can produce square artifacts, so we use a \emph{Random} sampling of the image and optimize multiple neighbourhoods until every pixel has converged. Our quantitative analysis shown of \tref{tab:optimization-order} shows that, while the \emph{Sliding} approach provides better results than independently fitting each pixel, the \emph{Random} approach provides the best results \new{overall}.


\paragraph{Results}

\begin{figure}[tb]
	\centering
	\includegraphics[width=\columnwidth]{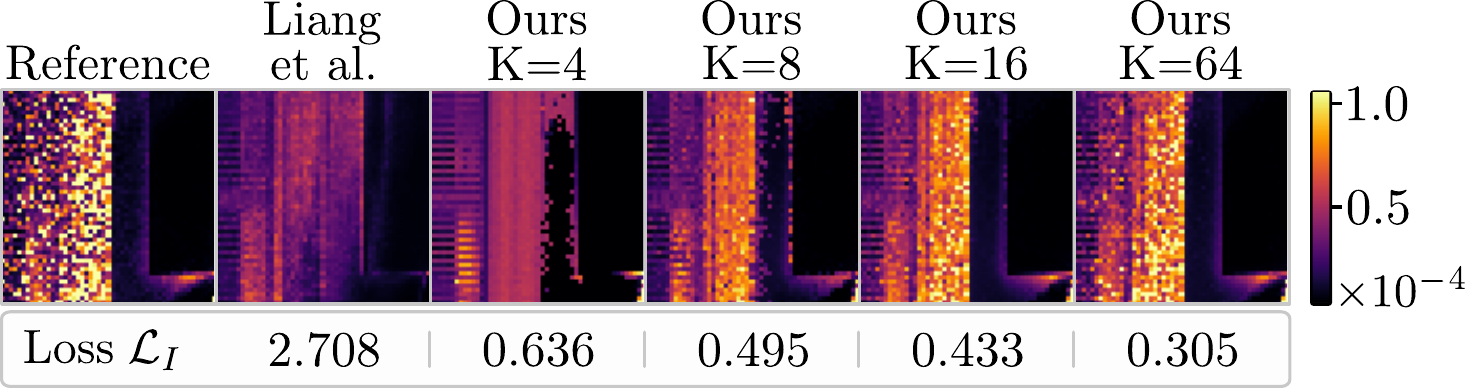}
	\caption{\new{Temporal slices and KLD loss metric $\mathcal{L}_I = \sum_\textbf{p} \mathcal{L}_{p}(I_\textbf{p}, I'_\textbf{p})$ in the \emph{Bathroom} scene \cite{MarcoSIGA2017DeepToF}, comparing each compressed image $I'$ with its original $I$ for different numbers $K$ of EMGs, yielding much better performance than previous work \cite{liang2020compression}.}}
	\label{fig:compare_k}
\end{figure}

\new{We evaluate the performance of signal compression for our \emph{Single pixel} (\eref{eq:single_pixel_optimization}) and \emph{Structure-aware} (\eref{eq:final_optimization_problem}) models, and compare to} the 3D convolutional autoencoder network recently proposed by Liang et al. \cite{liang2020compression}, designed for the same purposes.
\begin{figure}[tb]
	\centering
	\includegraphics[width=0.99\columnwidth]{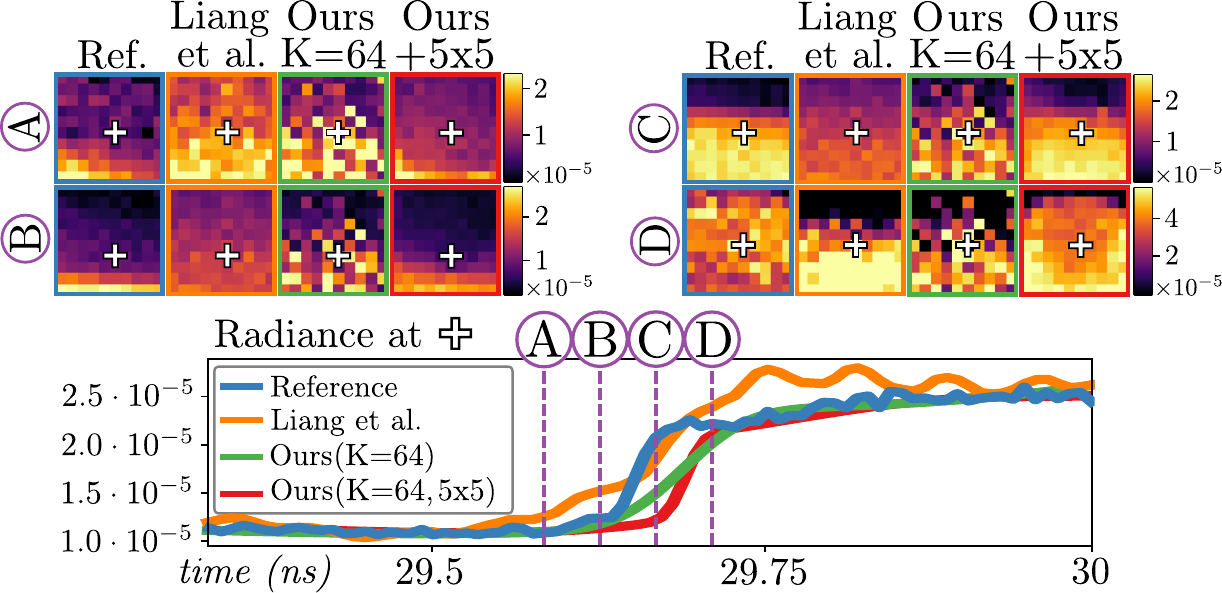}
	\caption{Signal representation of a 10x10 region of \emph{Staircase} \cite{galindo19-NLOSDataset}, as in \fref{fig:compare_decomp_k16}. The last row shows a spatial slice $I_\textbf{p}[t]$ for the pixel $\textbf{p}$ marked with a cross. The vertical dotted lines correspond to four temporal slices at different instants (A-D).}
	\label{fig:compare_decomp_k64_5x5}
\end{figure}
\begin{figure}[tb]
	\centering
	\includegraphics[width=0.99\columnwidth]{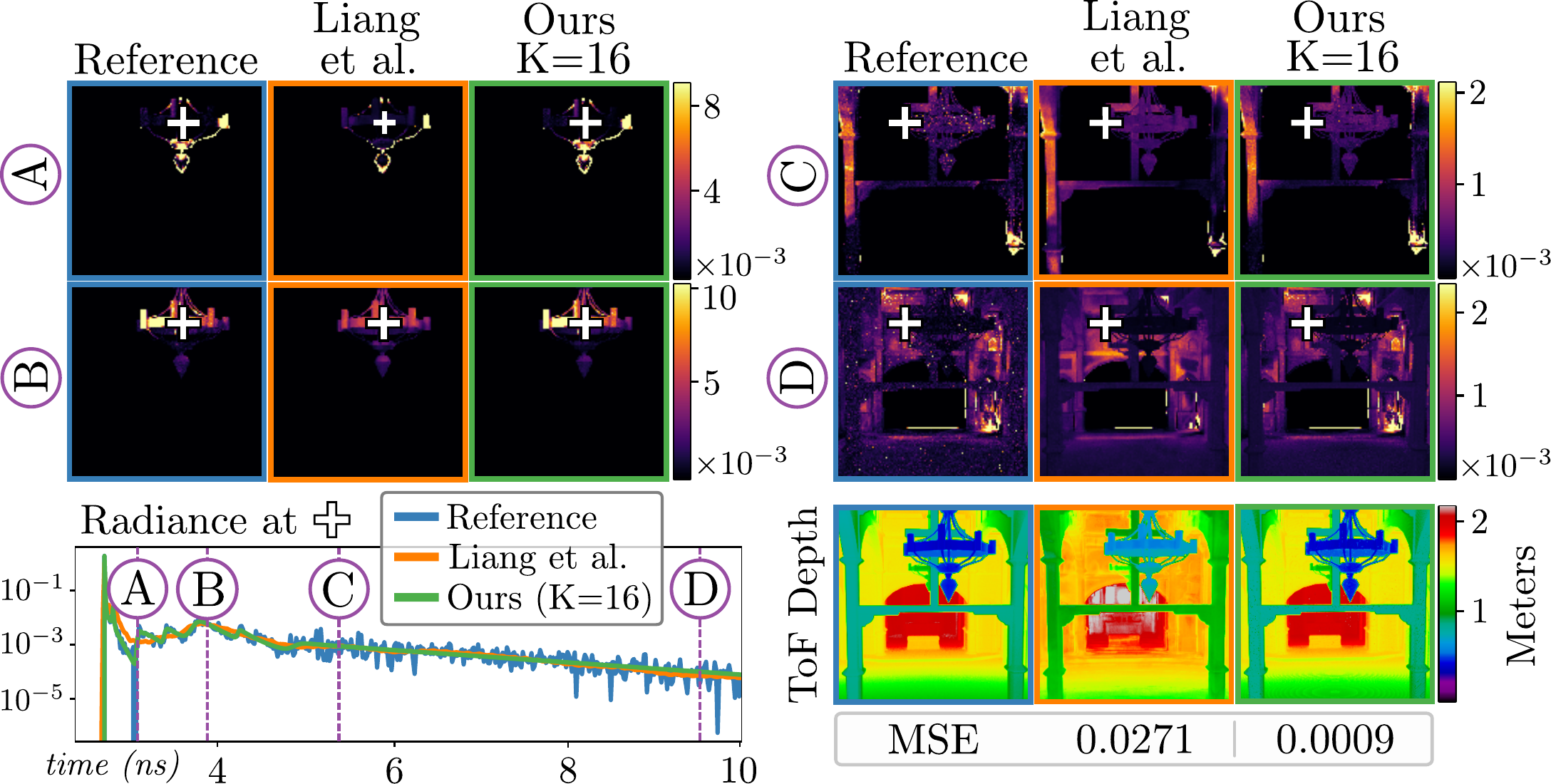}
 \caption{\new{Comparison on a LOS \emph{Church} scene \cite{MarcoSIGA2017DeepToF}. Our method using \emph{Single-pixel} mode (green) outperforms previous works (orange) in both signal reconstruction (spatial and temporal slices) and AMCW ToF depth estimation (bottom right)}.}
	\label{fig:compare_decomp_k16}
\end{figure}
\new{In our method, the number of EMG components $K$ introduces a trade-off between quality and efficiency. \fref{fig:compare_k} shows a comparison with the \emph{Single pixel} model. $K=4$ is enough to surpass deep-learning methods, increasing temporal similarity for higher $K$ values.
Execution times range from 5 seconds per pixel with $K=4$ EMGs, to 10 seconds with $K=64$ EMGs (Intel Xeon Gold 6140 CPU, using 10 threads).}
\fref{fig:compare_decomp_k64_5x5} uses $K=64$ \new{EMGs. It showcases the importance of the $5\times{5}$ window with spatial gradient constraints, added in the \emph{Strcuture-aware} model, compared to the \emph{Single pixel} model.}
%
\new{Following this, \fref{fig:nlos-reconstruction} shows} a real-world NLOS imaging application, where the capture setups produce noisier measurements when compared to simulated data. Our work provides a higher quality representation of both the signal and hidden geometry reconstruction, while also improving on the robustness to noise. The compression ratio is also higher, $T / (4\cdot K+2) \approx {42}$, \new {reducing the resulting file size from 365 MiB to 8.7 MiB.}
\new{Finally, \fref{fig:compare_decomp_k16} shows results on a LOS scene \cite{MarcoSIGA2017DeepToF} with $K=16$}, resulting in approximately 62 times less parameters than the original signal \new{with the \emph{Single pixel} model}. This almost doubles the compression ratio of the autoencoder network by Liang et al., \new{while providing more accurate representations, as shown in the spatial (A-D) and temporal (bottom left) slices of $I[i, j, t]$. Bottom-right images show a comparison on a practical application of AMCW ToF depth estimation, where our output (right) yields more consistent depth results than Liang et al.'s output (center) \cite{liang2020compression}.}
%


In conclusion, we presented a method to efficiently represent time-resolved light transport data based on exponentially modified Gaussians, significantly reducing the number of coefficients required to represent time-resolved transport. Our optimization loss, based on first-order spatial differences, preserves the spatial structure and reduces noise which are fundamental aspects in transient imaging applications. We demonstrated the benefits of our method in both LOS and NLOS scenarios, overcoming previous approaches targeted to transient light transport compression.
Relating the EMG components of our representation to higher-order illumination bounces may help to increase the quality of scene understanding applications such as multipath interference correction in depth imaging or hidden scene reconstruction. Additionally, analysis of the frequency components of our EMG-based representation could be exploited in wave-based NLOS imaging algorithms \cite{Lindell2019wave,liu2019non} to improve their computational efficiency. \new{Our implementation is public and can be found in the supplementary material.}


\begin{backmatter}


\bmsection{Funding} \emph{Left blank during review.}

\bmsection{Disclosures} The authors declare no conflicts of interest.

\bmsection{Data availability} Data underlying the results presented in this paper comes from the \emph{DeepToF} \cite{MarcoSIGA2017DeepToF} and \emph{Z-NLOS} \cite{galindo19-NLOSDataset} datasets.

\end{backmatter}

\bibliography{bibliography}

\begin{thebibliography}{10}
\newcommand{\enquote}[1]{``#1''}

\bibitem{liu2019non}
X.~Liu, I.~Guill{\'e}n, M.~La~Manna, J.~H. Nam, S.~A. Reza, T.~H. Le,
  A.~Jarabo, D.~Gutierrez, and A.~Velten, {\protect\JournalTitle{Nature}}
  (2019).

\bibitem{liang2020compression}
Y.~Liang, M.~Chen, Z.~Huang, D.~Gutierrez, A.~Mu{\~n}oz, and J.~Marco,
  {\protect\JournalTitle{Optics Letters}} \textbf{45}, 1986 (2020).

\bibitem{Lindell2019wave}
D.~B. Lindell, G.~Wetzstein, and M.~O'Toole, {\protect\JournalTitle{ACM Trans.
  Graph.}} \textbf{38}, 1 (2019).

\bibitem{Xin2019theory}
S.~Xin, S.~Nousias, K.~N. Kutulakos, A.~C. Sankaranarayanan, S.~G. Narasimhan,
  and I.~Gkioulekas, \enquote{A theory of {Fermat} paths for non-line-of-sight
  shape reconstruction,} in \emph{IEEE Computer Vision and Pattern Recognition
  (CVPR),}  (2019), pp. 6800--6809.

\bibitem{Heide2014}
F.~Heide, L.~Xiao, A.~Kolb, M.~B. Hullin, and W.~Heidrich,
  {\protect\JournalTitle{Opt. Express}} \textbf{22} (2014).

\bibitem{su2016material}
S.~Su, F.~Heide, R.~Swanson, J.~Klein, C.~Callenberg, M.~Hullin, and
  W.~Heidrich, \enquote{Material classification using raw time-of-flight
  measurements,} in \emph{Proceedings of the IEEE Conference on Computer Vision
  and Pattern Recognition,}  (2016), pp. 3503--3511.

\bibitem{jarabo2017recent}
A.~Jarabo, B.~Masia, J.~Marco, and D.~Gutierrez, {\protect\JournalTitle{Visual
  Informatics}} \textbf{1}, 65 (2017).

\bibitem{renna2020fast}
M.~Renna, J.~H. Nam, M.~Buttafava, F.~Villa, A.~Velten, and A.~Tosi,
  {\protect\JournalTitle{Instruments}} \textbf{4}, 14 (2020).

\bibitem{nam2021low}
J.~H. Nam, E.~Brandt, S.~Bauer, X.~Liu, M.~Renna, A.~Tosi, E.~Sifakis, and
  A.~Velten, {\protect\JournalTitle{Nature communications}} \textbf{12}, 1
  (2021).

\bibitem{marco2021NLOSvLTM}
J.~Marco, A.~Jarabo, J.~H. Nam, X.~Liu, M.~Ángel Cosculluela, A.~Velten, and
  D.~Gutierrez, \enquote{Virtual light transport matrices for non-line-of-sight
  imaging,} in \emph{2021 IEEE/CVF International Conference on Computer Vision
  (ICCV),}  (2021).

\bibitem{Peters2015}
C.~Peters, J.~Klein, M.~B. Hullin, and R.~Klein, {\protect\JournalTitle{ACM
  Trans. Graph.}} \textbf{34} (2015).

\bibitem{Kadambi2013}
A.~Kadambi, R.~Whyte, A.~Bhandari, L.~Streeter, C.~Barsi, A.~Dorrington, and
  R.~Raskar, {\protect\JournalTitle{ACM Trans. Graph.}} \textbf{32} (2013).

\bibitem{Heide2014diffuse}
F.~Heide, L.~Xiao, W.~Heidrich, and M.~B. Hullin, \enquote{Diffuse mirrors:
  {3D} reconstruction from diffuse indirect illumination using inexpensive
  time-of-flight sensors,} in \emph{IEEE Computer Vision and Pattern
  Recognition,}  (2014).

\bibitem{Lin2014}
J.~Lin, Y.~Liu, M.~B. Hullin, and Q.~Dai, \enquote{Fourier analysis on
  transient imaging with a multifrequency time-of-flight camera,} in \emph{IEEE
  Computer Vision and Pattern Recognition,}  (2014).

\bibitem{Wu2014}
D.~Wu, A.~Velten, M.~O'Toole, B.~Masia, A.~Agrawal, Q.~Dai, and R.~Raskar,
  {\protect\JournalTitle{International Journal of Computer Vision}}
  \textbf{107} (2014).

\bibitem{MarcoSIGA2017DeepToF}
J.~Marco, Q.~Hernandez, A.~Mu\~{n}oz, Y.~Dong, A.~Jarabo, M.~Kim, X.~Tong, and
  D.~Gutierrez, {\protect\JournalTitle{ACM Transactions on Graphics (SIGGRAPH
  Asia 2017)}} \textbf{36} (2017).

\bibitem{galindo19-NLOSDataset}
M.~Galindo, J.~Marco, M.~O'Toole, G.~Wetzstein, D.~Gutierrez, and A.~Jarabo,
  \enquote{A dataset for benchmarking time-resolved non-line-of-sight imaging,}
   (2019).

\bibitem{kullback1951information}
S.~Kullback and R.~A. Leibler, {\protect\JournalTitle{The annals of
  mathematical statistics}} \textbf{22}, 79 (1951).

\end{thebibliography}

\bibliographyfullrefs{bibliography}


\ifthenelse{\equal{\journalref}{aop}}{%
\section*{Author Biographies}
\begingroup
\setlength\intextsep{0pt}
\begin{minipage}[t][6.3cm][t]{1.0\textwidth} 
  \begin{wrapfigure}{L}{0.25\textwidth}
    \includegraphics[width=0.25\textwidth]{john_smith.eps}
  \end{wrapfigure}
  \noindent
  {\bfseries John Smith} received his BSc (Mathematics) in 2000 from The University of Maryland. His research interests include lasers and optics.
\end{minipage}
\begin{minipage}{1.0\textwidth}
  \begin{wrapfigure}{L}{0.25\textwidth}
    \includegraphics[width=0.25\textwidth]{alice_smith.eps}
  \end{wrapfigure}
  \noindent
  {\bfseries Alice Smith} also received her BSc (Mathematics) in 2000 from The University of Maryland. Her research interests also include lasers and optics.
\end{minipage}
\endgroup
}{}

\end{document}